%% file: 0.main.tex
\documentclass[sigconf,natbib=true,anonymous=false]{acmart}

\usepackage{booktabs} 
\usepackage{algorithm}
\usepackage{algorithmic} 

\usepackage{caption}      
\usepackage{subcaption}   
\usepackage{ifthen}
\usepackage{amsmath} 
\usepackage{latexsym}
\usepackage{CJK}

\usepackage{amsfonts,amssymb}
\usepackage{ifthen}
\usepackage{graphicx}
\usepackage{float}
\usepackage{multirow}
\usepackage{balance}
\usepackage{graphics}
\usepackage{enumerate}
\usepackage{verbatim}
\usepackage{float}
\usepackage{booktabs}
\usepackage[hang,flushmargin]{footmisc}
\usepackage{bbold}
\usepackage{color}
\usepackage{xspace}
\usepackage{rotating}
\usepackage{enumitem}
\usepackage[table]{xcolor} 

\newcommand{\paratitle}[1]{\vspace{1.5ex}\noindent\textbf{#1}}

\newcommand{\ignore}[1]{}

\newcommand{\baby}{\textsc{INFNet}\xspace}

\renewcommand\footnotetextcopyrightpermission[1]{}
\AtBeginDocument{%
  }

\setcopyright{acmlicensed}
\copyrightyear{2018}
\acmYear{2018}
\acmDOI{XXXXXXX.XXXXXXX}
\acmConference[Conference acronym 'XX]{Make sure to enter the correct
  conference title from your rights confirmation email}{June 03--05,
  2018}{Woodstock, NY}
\acmISBN{978-1-4503-XXXX-X/18/06}

\begin{document}
\title{Aggregate and Broadcast: Scalable and Efficient Feature Interaction for Recommender Systems}

\author{Kaiyuan Li}
\affiliation{
  \institution{Kuaishou Technology, Beijing, China}
  \city{}
  \country{}
}
\email{likaiyuan03@kuaishou.com}


\author{Yongxiang Tang}
\affiliation{
  \institution{Kuaishou Technology, Beijing, China}
  \city{}
  \country{}
}
\email{tangyongxiang@kuaishou.com}

\author{Wenzheng Shu}
\affiliation{
  \institution{Kuaishou Technology, Beijing, China}
  \city{}
  \country{}
}
\email{shuwenzheng@kuaishou.com}

\author{Yanxiang Zeng}
\affiliation{
  \institution{Kuaishou Technology, Beijing, China}
  \city{}
  \country{}
}
\email{zengyanxiang@kuaishou.com}

\author{Chao Wang}
\affiliation{
  \institution{Kuaishou Technology, Beijing, China}
  \city{}
  \country{}
}
\email{wangchao32@kuaishou.com}

\author{Yanhua Cheng}
\affiliation{
  \institution{Kuaishou Technology, Beijing, China}
  \city{}
  \country{}
}
\email{chengyanhua@kuaishou.com}

\author{Xialong Liu}
\affiliation{
  \institution{Kuaishou Technology, Beijing, China}
  \city{}
  \country{}
}
\email{liuxialong2007@sina.com}

\author{Peng Jiang}
\affiliation{%
  \institution{Kuaishou Technology, Beijing, China}
  \city{}
  \country{}
}
\email{jp2006@139.com}

\renewcommand{\shortauthors}{Kaiyuan Li et al.}

\begin{abstract}
Feature interaction is a core ingredient in ranking models for large-scale recommender systems, yet making it both expressive and efficiently scalable remains challenging. Exhaustive pairwise interaction is powerful but incurs quadratic complexity in the number of tokens/features, while many efficient alternatives rely on restrictive structures that limit information exchange. We further identify two common bottlenecks in practice: (1) early aggregation of behavior sequences compresses fine-grained signals, making it difficult for deeper layers to reuse item-level details; and (2) late fusion injects task signals only at the end, preventing task objectives from directly guiding the interaction process.

To address these issues, we propose the Information Flow Network (INFNet), a lightweight architecture that enables scalable, task-aware feature interaction with linear complexity. INFNet represents categorical features, behavior sequences, and task identifiers as tokens, and introduces a small set of hub tokens for each group to serve as communication hubs.
Interaction is realized through an efficient aggregate-and-broadcast information flow: hub tokens aggregate global context across groups via cross-attention, and a lightweight gated broadcast unit injects the refined context back to update the categorical, sequence, and task tokens. This design supports width-preserving stacking that preserves item-level signals in sequence and enables task-guided interaction throughout the network, while reducing interaction cost from quadratic to linear in the number of feature tokens.

Experiments on a public benchmark and a large-scale industrial dataset demonstrate that INFNet consistently outperforms strong baselines and exhibits strong scaling behavior. In a commercial online advertising system, deploying INFNet improves revenue by +1.587\% and click-through rate by +1.155\%.
\end{abstract}



\maketitle
\section{Introduction}
\input{1.intro}

\section{Related Work}
\input{2.related}

\section{Methodology}
\input{3.method}

\section{Experiment}
\input{4.exp}

\section{Conclusion}
\input{5.con}

\balance

\bibliographystyle{ACM-Reference-Format}
\bibliography{reference}

\end{document}

%% file: 1.intro.tex
\begin{figure*}[h]
    \centering
    \includegraphics[width=1\linewidth]{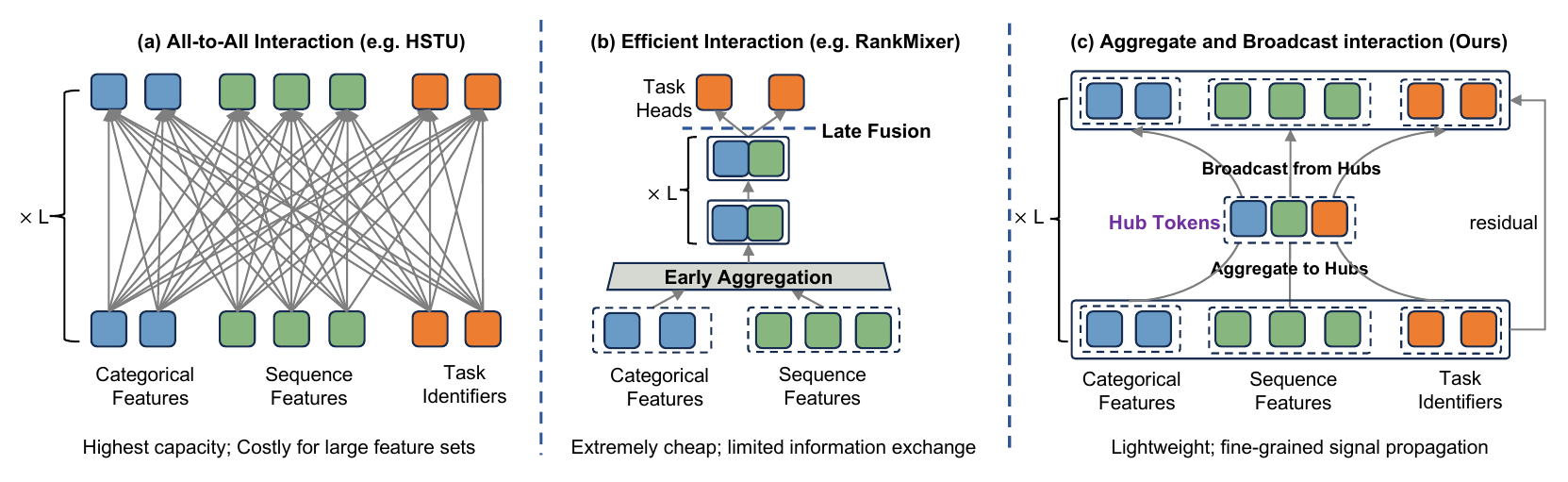}
    \caption{
        \textbf{Comparison of Feature Interaction Paradigms when stacking L layers.} 
        (a) \textbf{All-to-All Interaction} (e.g., HSTU) enables exhaustive connectivity but suffers from quadratic complexity. 
        (b) \textbf{Efficient Interaction} (e.g., RankMixer) relies on \textit{Early Aggregation} to compress sequences and \textit{Late Task Fusion}, which acts as a bottleneck for information flow. 
        (c) \textbf{Aggregate-and-Broadcast Interaction (Ours)}: INFNet utilizes an Aggregate-and-Broadcast mechanism mediated by hub tokens. It maintains a width-preserving architecture with linear complexity, ensuring effective signal propagation and task-guided interaction.
    }
    \label{fig:paradigm_comparison}
\end{figure*}

Feature interaction serves as the backbone of modern ranking models in large-scale recommender systems~\cite{zhang2023reformulating,zhang2023fibinet++,gui2023hiformer,li2024dcnv3,tian2023ufin,sang2024feature}, especially in content discovery and short-video recommendation. To accurately predict user responses, the system must analyze intricate dependencies among diverse features, for example, the correlation between a user's viewing history and video attributes under specific contexts. By effectively modeling these dependencies, ranking models can better predict multi-objective user feedback, such as playback completion, likes, and clicks.

With the development of deep learning, the field has evolved from shallow models such as Factorization Machines\cite{rendle2010factorization, lian2018xdeepfm, guo2017deepfm} to sophisticated attention-based architectures\cite{song2019autoint,zhai2024actions,xu2025climber,zhang2025onetrans}. Ideally, a ranking model should allow every feature to interact broadly with others to capture complex, high-order dependencies. However, as illustrated in Figure~\ref{fig:paradigm_comparison}(a), approaches based on all-to-all interaction (e.g., AutoInt\cite{song2019autoint}, HSTU\cite{zhai2024actions}) provide rich connectivity but often incur quadratic cost with respect to the number of input tokens, where each token corresponds to a categorical feature or a behavior element.
However, in large-scale industrial scenarios, efficiency is as critical as expressiveness. Online serving systems must meet strict latency limits (often below 30ms per request) while handling hundreds of feature fields and multi behavior sequences. This creates a sharp conflict: exhaustive interaction is often prohibitively expensive in practice.
Consequently, recent work has increasingly focused on lightweight interaction designs. Representative approaches such as \textbf{RankMixer}~\cite{zhu2025rankmixer} and \textbf{OneTrans}~\cite{zhang2025onetrans} adopt interaction patterns whose computational cost scales linearly or near-linearly with the number of input tokens. By leveraging MLP-based mixing or efficient transformer-style aggregation, these methods achieve favorable inference latency and scalability in real-world systems.

While these lightweight designs excel in efficiency and practical scalability, scaling the model does not always translate into stronger interaction capacity, partly due to structural choices that restrict how information is preserved and exchanged during interaction. From an information-flow perspective, two commonly adopted design patterns warrant a rethinking.

First, For sequential features, efficiency-oriented models
 \cite{zhou2018deep, feng2019deep, zhu2025rankmixer} typically employ \textbf{early aggregation} (e.g., Sum-Pooling or target-attention) to compress variable-length behavior sequences into fixed-length vectors before the interaction stage. While computationally efficient, this operation acts as a bandwidth constriction at the source. By projecting complex temporal sequences into a lower-dimensional representation, the model may reduce the fidelity of fine-grained signals. As a result, even with larger downstream capacity, interaction layers may have limited access to the original temporal details, weakening token-level dependencies between specific past events and the target item.

Second, regarding task modeling, many multi-objective architectures \cite{wang2024home,su2024stem,zhang2024m3oe,zhu2025rankmixer,zhang2025onetrans} adopt a sequential design where task signals are fused only at the final output stage (\textbf{late fusion}). This design often leads to a relatively task-agnostic interaction process. Without early access to task-specific context, the feature interaction module produces generic representations. While often sufficient for simpler settings or single-task objectives, this separation can constrain specialization for distinct ranking objectives, potentially under-utilizing model capacity when scaling up in multi-objective scenarios.

The detailed illustration is provided in Figure~\ref{fig:paradigm_comparison}(b). To address these limitations while preserving linear-time efficiency, we propose the \textbf{Information Flow Network (INFNet)}, which reconceptualizes feature interaction as a structured \emph{aggregate-and-broadcast} information flow (see Figure~\ref{fig:paradigm_comparison}(c)). Rather than relying on early compression or task-agnostic interaction, INFNet explicitly models how information is collected, transformed, and redistributed across heterogeneous feature groups.

INFNet explicitly organizes features into three distinct groups: categorical features, behavior sequences, and task identifiers. Specifically, we treat each categorical feature as a categorical token, every individual item in behavior history as a sequence token, and each target objective as a task token. These uncompressed tokens are maintained in a width-preserving stacked architecture. To reconcile the computational cost with strict latency requirements, we assign a small set of hub tokens to each feature group. The architecture orchestrates interaction through a symmetric mechanism consisting of two alternating phases:

(1) An \textbf{Aggregation Flow}, where each group's hub tokens facilitate interaction by acting as queries to selectively aggregate global context among feature groups via cross-attention;

(2) A \textbf{Broadcast Flow}, where the aggregated information is then distributed via a intra-group broadcast gated unit to update the group-specific tokens.

Crucially, through this group-wise, hub-token-mediated mechanism, INFNet ensures that feature signals and task signals are efficiently exchanged across domains. This design achieves a favorable trade-off: it preserves the linear computational complexity characteristic of lightweight models while improving scaling capacity compared with strong lightweight baselines.

The main contributions of this work are summarized as follows:

\begin{itemize}[leftmargin=*]
    \item We analyze efficient ranking models through the lens of information flow, pinpointing "early sequence aggregation" and "late task fusion" as bottlenecks that limit scaling under efficiency constraints in lightweight architectures.
    \item We propose INFNet, utilizing an aggregate-and-broadcast mechanism mediated by hub tokens. It achieves width-preserving and task-guided interaction with linear complexity.
    \item We demonstrate that INFNet exhibits superior scaling capabilities. Compared with strong lightweight baselines, it continues to gain performance as model size increases.
    \item Extensive offline experiments and online A/B testing in a large-scale commercial advertising system confirm INFNet's effectiveness, delivering substantial uplifts of +1.587\% in revenue and +1.155\% in CTR.
\end{itemize}

%% file: 2.related.tex
Industrial-grade deep learning recommendation models (DLRMs) typically integrate feature interaction, user behavior modeling, and multi-task learning within a unified ranking stack~\cite{zhu2025rankmixer, zhang2025onetrans}. Feature interaction has progressed from factorization-based foundations such as FM~\cite{rendle2010factorization} and its variant extensions~\cite{juan2016field, pan2018field, guo2017deepfm} to high-order structural designs including DCN-V2~\cite{wang2021dcn}, PNN~\cite{qu2016product}, and attention-driven architectures like AutoInt~\cite{song2019autoint} and FiBiNET~\cite{huang2019fibinet}. Recent refinements such as FinalMLP~\cite{mao2023finalmlp}, xDeepInt~\cite{yan2023xdeepint} and FuXi-$\alpha$~\cite{ye2025fuxi}, allow for controllable capacity allocation across fields, improving scalability.

\begin{figure*}[h]
    \centering
    \includegraphics[width=1.0\linewidth]{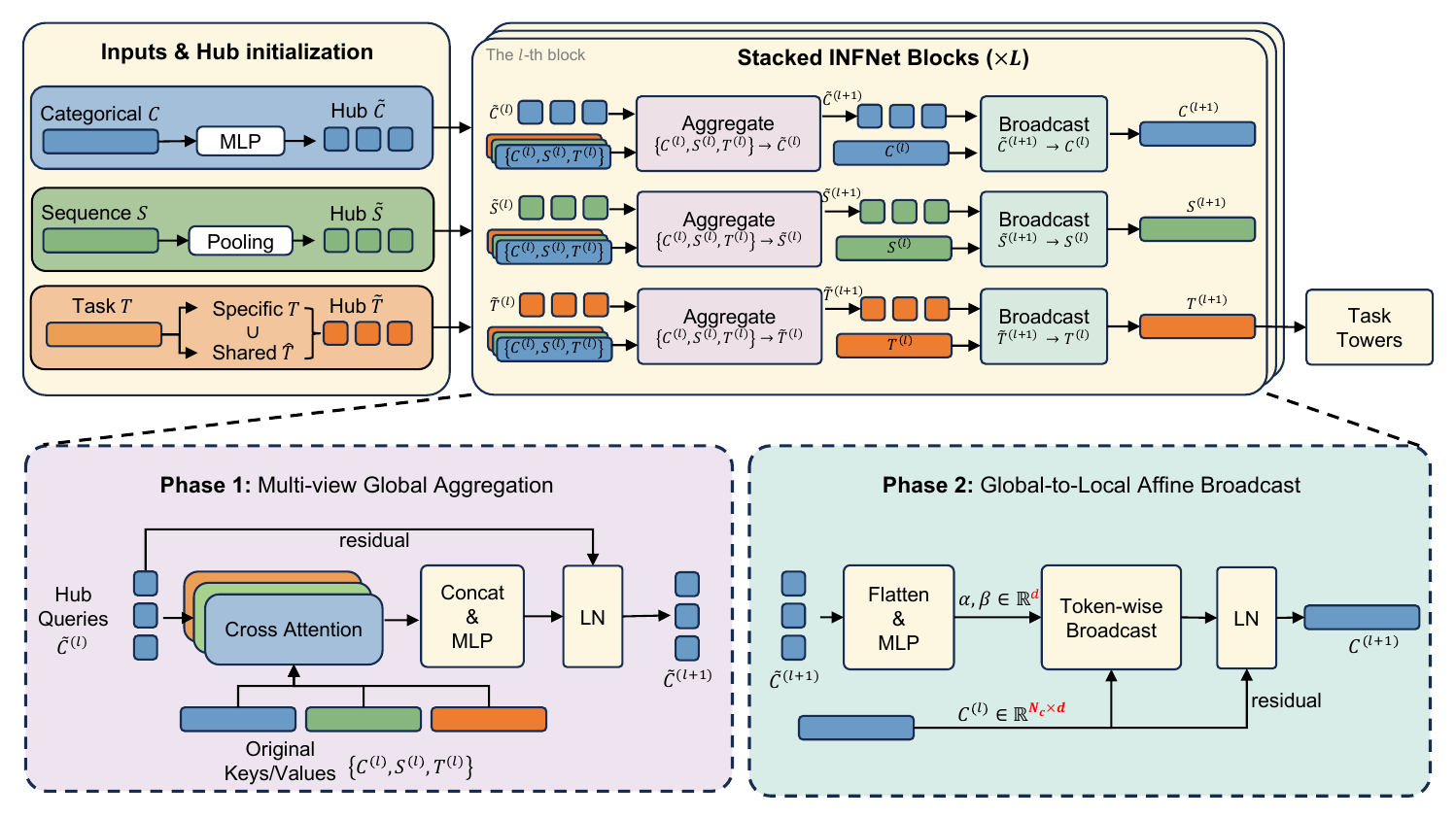}
    \caption{\small{The overall architecture of INFNet. 
    \textbf{(Top Panel)} The model workflow. \textit{Left}: Group-wise hubs are initialized via distinct strategies (MLP for Categorical, Pooling for Sequence, and Hybrid for Task). \textit{Right}: The stacked INFNet Blocks process all feature groups in the aggregation-and-broadcast mechanism.
    \textbf{(Bottom Panel)} Detailed illustration of the two-phase interaction mechanism, \textbf{using the Categorical Group as an example}:
    \textbf{1) Multi-view Global Aggregation:} Hubs (e.g., $\tilde{C}^{(l)}$) act as queries to harvest global context from the union of all original tokens ($\{C, S, T\}$) via Cross-Attention.
    \textbf{2) Global-to-Local Affine Broadcast:} The refined hubs ($\tilde{C}^{(l+1)}$) generate affine parameters ($\alpha, \beta$) to modulate their corresponding original tokens ($C^{(l)}$), effectively broadcasting global context back to local features.}}
    \label{fig:model}
\end{figure*}

Simultaneously, behavior modeling has shifted from simple pooling to sophisticated attention mechanisms in DIN~\cite{zhou2018deep} and DIEN~\cite{zhou2019deep}, and further to Transformer-based BST~\cite{chen2019behavior}. To handle the computational overhead of lifelong sequences, strategies such as SIM~\cite{pi2020search}, ETA~\cite{chen2021end}, TWIN/TWIN-V2~\cite{chang2023twin, si2024twin} have been widely deployed.

Despite these advancements, a fundamental tension persists between structural unification and scaling efficiency. On one hand, recent unified architectures like InterFormer~\cite{zeng2024interformer}, OneTrans~\cite{zhang2025onetrans}, and Uni-CTR~\cite{fu2025unified} aim to model categorical and sequential features jointly to eliminate information isolation. On the other hand, the emergence of Scaling Laws in recommendation pioneered by heavy-duty architectures like HSTU~\cite{zhai2024actions}, and Climber~\cite{xu2025climber} highlights the necessity of predictable capacity growth. Efficient scaling mechanisms, such as WuKong~\cite{zhang2024wukong} and RankMixer~\cite{zhu2025rankmixer}, seek to reconcile large-scale modeling with industrial efficiency. However, unified frameworks often suffer from quadratic complexity, while efficiency-centric models often rely on coarse-grained early aggregation or late fusion, which prematurely collapses fine-grained signals.

Multi-objective ranking further compounds this complexity~\cite{wang2024home, zhang2024m3oe}. Beyond foundational frameworks like MMoE~\cite{ma2018modeling} and PLE~\cite{tang2020progressive}, recent works such as HiFormer~\cite{gui2023hiformer}, STEM~\cite{su2024stem}, and AITM~\cite{xi2021modeling} introduce hierarchical task-aware routing and adaptive information transfer. Despite these efforts, many deployed interaction backbones remain task-agnostic, with task-specific signals injected only at late stages~\cite{zhu2025rankmixer, zhang2024wukong}.

Seeking to resolve the quadratic complexity of exhaustive interaction, research in NLP and CV has explored latent attention mechanisms. Set Transformer~\cite{lee2019set} introduced inducing points to reduce complexity, while the Perceiver ~\cite{jaegle2021perceiver} generalized this into a latent-token-based cross-attention paradigm. While achieving near-linear scaling, these methods are primarily optimized for homogeneous data and lack task sensitivity. 

In this paper, we explore a unified approach to sequential modeling, feature interaction, and multi-task learning. By leveraging latent attention mechanisms, we aim to reduce interaction complexity and enhance the overall efficiency and scalability of the model.

%% file: 3.method.tex
In this section, we present the \textbf{Information Flow Network (INFNet)}, an aggregate and broadcast ranking architecture designed to reconcile the conflict between serving efficiency and deep interaction capacity. As illustrated in Figure~\ref{fig:model}, unlike traditional methods that either compress features early or perform expensive all-to-all interactions, INFNet reconceptualizes feature interaction as a structured information flow mediated by a compact set of Hub Tokens.

We first describe a group-wise tokenization strategy that explicitly accounts for the scale, structure, and semantics of heterogeneous feature types. Next, we describe the core Aggregate-and-Broadcast mechanism within the stackable INFNet block. Finally, we present the multi-task optimization objective. Table~\ref{tab:notation} summarizes the key notations.

\begin{table}[h]
    \centering
    \footnotesize
    \caption{Summary of Key Notations.}
    \label{tab:notation}
    \setlength{\tabcolsep}{2.5pt}
    \renewcommand{\arraystretch}{1.2}
    \begin{tabular}{l|l}
    \hline
    \textbf{Symbol} & \textbf{Description} \\
    \hline
    $N_c, N_t$ & Number of Categorical fields and Tasks \\
    $N_s$ & Total sequence length \\
    $N_{\mathrm{in}}, N_{\mathrm{hub}}$ & Total number of Original Tokens and Hub Tokens \\
    $d, L$ & Feature dimension; Number of INFNet blocks \\
    \hline
    $\mathbf{X}_g, \tilde{\mathbf{X}}_g$ & Original \& Hub tokens for group $g \in \{C, S, T\}$ \\
    $n_c, n_s$ & Number of virtual hubs for Categorical / Sequence \\
    $n_{\mathrm{shared}}$ & Number of shared hubs for Task group \\
    \hline
    $\operatorname{CA}(\cdot)$ & Cross-Attention operator \\
    $\operatorname{LN}(\cdot)$ & Layer Normalization operator \\
    $\operatorname{BGU}(\cdot)$ & Broadcast Gated Unit operator \\
    $\boldsymbol{\alpha}, \boldsymbol{\beta}$ & Scaling and Shifting parameters in BGU \\
    \hline
    \end{tabular}
\end{table}

\subsection{Group-wise Tokenization \& Hub Initialization}
To enable fine-grained interaction without introducing a structural bottleneck, we unify all inputs into three distinct groups: \textit{Categorical Features}, \textit{Behavior Sequences}, and \textit{Task Identifiers}. For each group, we maintain two sets of representations: \textbf{Original Tokens} ($\mathbf{X}$), which preserve token-level granularity (avoiding pre-interaction pooling), and \textbf{Hub Tokens} ($\tilde{\mathbf{X}}$), which serve as efficient communication nodes. Crucially, the definition and initialization of these hubs are tailored to the scale and semantics of each group. We analyze the impact of different hub token settings in Section~\ref{sec:ablation}.

\subsubsection{Categorical Features}
This group encompasses discrete attributes (e.g., User ID, Item ID) and continuous dense features (e.g., age, view counts). To ensure unified processing, we first discretize all dense features into buckets via logarithmic or quantile binning, treating the bucket indices as categorical IDs. 
For $N_c$ resulting features, we map them to an original token matrix $\mathbf{C} \in \mathbb{R}^{N_c \times d}$, where $d$ is the embedding dimension.

\noindent \textbf{Hub Generation (Compression Strategy).}
Since the number of categorical features $N_c$ is typically large (often hundreds or thousands in practice), performing pairwise interaction directly on $\mathbf{C}$ would be computationally prohibitive. Therefore, we aim to capture a compressed global view of the static context. Instead of random initialization, we generate $n_c$ virtual hubs ($n_c \ll N_c$) by projecting the flattened original features. Formally, let $\phi^{\mathrm{cat}}(\cdot)$ denote a Multi-Layer Perceptron (MLP):
\begin{equation}
\tilde{\mathbf{C}} = \operatorname{Reshape}\Bigl( \phi^{\mathrm{cat}}[\operatorname{Flatten}(\mathbf{C})] \Bigr) \in \mathbb{R}^{n_c \times d}.
\end{equation}
This initialization strategy semantically grounds the hubs with global feature co-occurrence patterns, providing a robust starting point for subsequent interactions.

\subsubsection{Behavior Sequences}
User behaviors are naturally heterogeneous, consisting of various actions such as clicks, likes, and purchases. We organize the user history into $B$ distinct behavior types, denoted as $\{\mathcal{B}_1, \dots, \mathcal{B}_B\}$.
Within the $b$-th type, the sequence consists of $T_b$ behavior items, strictly sorted in chronological order. Unlike methods that pool these sequences early, we treat every individual item together with its side information as a distinct Original Token to preserve item-level temporal evidence.
For the $t$-th item in behavior type $b$ with raw embedding $\mathbf{e}_{b,t}$, we inject temporal dynamics via learnable absolute position $\mathbf{p}_t \in \mathbb{R}^d$ and relative time-interval embeddings $\mathbf{r}_t \in \mathbb{R}^d$:
\begin{equation}
\mathbf{s}_{b,t} = \mathbf{e}_{b,t} + \mathbf{p}_t + \mathbf{r}_t.
\end{equation}
Concatenating all items across all types yields the full original sequence matrix $\mathbf{S} \in \mathbb{R}^{N_s \times d}$, where $N_s = \sum_{b=1}^{B} T_b$.

\noindent \textbf{Hub Generation (Type-Specific Aggregation).}
To serve as communication hubs for this group, we assign exactly one virtual hub to each behavior type, resulting in $n_s = B$ hubs. 
Crucially, different behavior types may reside in different semantic spaces. To capture this diversity, we employ distinct, type-specific projection networks. For the $b$-th behavior type, all items share a specific mapping $\operatorname{MLP}_b^{\mathrm{seq}}(\cdot)$. The hub $\tilde{\mathbf{s}}_b$ is obtained by averaging projected features within that type:
\begin{equation}
\tilde{\mathbf{s}}_b = \frac{1}{T_b} \sum_{t=1}^{T_b} \operatorname{MLP}_b^{\mathrm{seq}}(\mathbf{s}_{b,t}).
\end{equation}
Collecting these yields $\tilde{\mathbf{S}} = [\tilde{\mathbf{s}}_1; \dots; \tilde{\mathbf{s}}_B] \in \mathbb{R}^{B \times d}$. This design allows each hub to function as a specialized ``Interest Prototype,'' summarizing the user's preference in a specific domain.

\subsubsection{Task Identifiers}
We explicitly model task objectives as inputs to guide the feature interaction. Unlike the massive feature groups described above, the number of tasks $N_t$ is small.
For this group, we initialize $N_t$ learnable vectors as the original Tokens, denoted as $\mathbf{T} = [\mathbf{t}_1; \dots; \mathbf{t}_{N_t}] \in \mathbb{R}^{N_t \times d}$. Each token $\mathbf{t}_i$ semantically represents a specific optimization objective (e.g., click or like) and serves as the anchor for the final task-specific prediction head.

\noindent \textbf{Hub Generation (Shared-Specific Augmentation).}
Unlike the categorical and sequence groups where hubs are compressed from originals, the task group requires increased capacity to model commonalities between tasks. Therefore, instead of compression, we adopt a shared-specific augmentation strategy to construct the hubs.
We introduce $n_{\mathrm{shared}}$ additional learnable tokens as shared hubs $\tilde{\mathbf{T}}_{\mathrm{shared}} \in \mathbb{R}^{n_{\mathrm{shared}} \times d}$ to capture implicit common knowledge. The final hub tokens for the task group are constructed by concatenating the original specific tokens with these shared tokens:
\begin{equation}
    \tilde{\mathbf{T}} = \operatorname{Concat}[\mathbf{T}; \tilde{\mathbf{T}}_{\mathrm{shared}}] \in \mathbb{R}^{(N_t + n_{\mathrm{shared}}) \times d}.
\end{equation}
By utilizing this augmented set as hubs, INFNet ensures that the interaction is guided by both specific task requirements (via $\mathbf{T}$) and generalized knowledge (via $\tilde{\mathbf{T}}_{\mathrm{shared}}$).

\subsection{The INFNet Block}
The core of our architecture is the stackable \textbf{INFNet Block}. Let $l \in \{0,\dots,L-1\}$ index the blocks. 
Given the total input tokens $N_{\mathrm{in}} = N_c + N_s + N_t$, our goal is to achieve an interaction cost that scales linearly with $N_{\mathrm{in}}$ under a fixed hub budget. 
To this end, we design a symmetric \textbf{Aggregate-and-Broadcast} mechanism. Each block operates in two strictly alternating phases, ensuring that global context is first aggregated into a small set of hubs and then broadcast to refine local tokens.

\subsubsection{Phase 1: Multi-View Global Aggregation}
In this phase, the designated hub tokens act as queries to harvest comprehensive context from the entire feature space. The goal is to update the hubs' latent belief by absorbing signals from both their own domain (to preserve semantic identity) and external domains (to capture cross-view correlations) in a computationally efficient manner with linear complexity.
Given the symmetric nature of our architecture across feature groups, we take the categorical group as a representative instance to detail the aggregation mechanism, while other groups (sequence and task) follow an identical protocol.

We first define the standard Cross-Attention (CA) operator, which retrieves information from a Key-Value pair $(\mathbf{K}, \mathbf{V})$ based on a Query $\mathbf{Q}$:
\begin{equation}
\operatorname{CA}(\mathbf{Q},\mathbf{K},\mathbf{V}) = \operatorname{Softmax}\left(\frac{\mathbf{Q}\mathbf{K}^\top}{\sqrt{d}}\right)\mathbf{V}.
\end{equation}

To capture a holistic view, the categorical hubs $\tilde{\mathbf{C}}^{(l)}$ independently query three distinct sources: the intrinsic categorical tokens (self-view), the behavior sequence tokens (temporal-view), and the task tokens (goal-view). This yields three context vectors:
\begin{align}
    \mathbf{Z}_{C} &= \operatorname{CA}(\tilde{\mathbf{C}}^{(l)}, \mathbf{C}^{(l)}, \mathbf{C}^{(l)}), \nonumber \\
    \mathbf{Z}_{S} &= \operatorname{CA}(\tilde{\mathbf{C}}^{(l)}, \mathbf{S}^{(l)}, \mathbf{S}^{(l)}), \nonumber \\
    \mathbf{Z}_{T} &= \operatorname{CA}(\tilde{\mathbf{C}}^{(l)}, \mathbf{T}^{(l)}, \mathbf{T}^{(l)}).
\end{align}
To aggregate different views, we employ a concatenation-based fusion strategy. The retrieved contexts are concatenated and fused via a learnable linear projection to synthesize the updated information, which is then added to the residual hub stream:
\begin{align}
    \mathbf{H}_{\mathrm{fused}} &= \operatorname{Linear}\left( \operatorname{Concat}[\mathbf{Z}_{C}; \mathbf{Z}_{S}; \mathbf{Z}_{T}] \right), \nonumber \\
    \tilde{\mathbf{C}}^{(l+1)} &= \operatorname{LN}\left(\tilde{\mathbf{C}}^{(l)} + \mathbf{H}_{\mathrm{fused}}\right).
\end{align}
Following this unified protocol, the Sequence hubs $\tilde{\mathbf{S}}$ and Task hubs $\mathbf{T}$ (including both shared and specific tokens) symmetrically perform the same aggregation process, ensuring that every hub evolves by integrating multi-view evidence relevant to its semantic role. Note that $\operatorname{LN}(\cdot)$ denotes Layer Normalization.

\noindent \textbf{Complexity Analysis.} 
Let $N_{\mathrm{in}}$ denote the total number of original input tokens and $N_{\mathrm{hub}}$ be the total number of hub tokens. In standard self-attention mechanisms, modeling global interactions requires calculating an affinity matrix of size $N_{\mathrm{in}} \times N_{\mathrm{in}}$, leading to a quadratic complexity of $O(N_{\mathrm{in}}^2 d)$. 
In contrast, our Phase 1 aggregation restricts the queries to the compact hub set. The cross-attention operations are performed between $N_{\mathrm{hub}}$ queries and $N_{\mathrm{in}}$ keys/values. Consequently, the computational cost is reduced to $O(N_{\mathrm{hub}} N_{\mathrm{in}} d)$. Since the number of hubs is a small constant tailored to feature fields (i.e., $N_{\mathrm{hub}} \ll N_{\mathrm{in}}$), our method achieves \textbf{linear complexity} $O(N_{\mathrm{in}})$ with respect to the input scale, significantly enhancing scalability.

\subsubsection{Phase 2: Global-to-Local Affine Broadcast}
Once the hubs have synthesized the multi-view global context, the \textit{Broadcast Flow} disseminates this high-level information back to fine-grained local tokens. We propose a unified mechanism, the Broadcast Gated Unit (BGU), to perform this operation efficiently across all feature groups (Categorical, Sequence, and Task).

\noindent \textbf{The BGU Mechanism.}
The BGU is designed to modulate the original tokens by injecting global context. 
Functionally, it employs an affine transformation logic (scaling and shifting). 
While this calculation mathematically aligns with the Feature-wise Linear Modulation (FiLM) mechanism~\cite{perez2018film}, originally proposed for visual reasoning, our core innovation lies in repurposing it for the Hub-to-Leaf broadcasting context.
Instead of using FiLM for multi-modal fusion, we leverage its parameter efficiency to bridge the dimensionality gap between the compressed global hubs and the massive local tokens.
Unlike simple multiplicative gating which can only attenuate features, this affine design enables the model to highlight important features ($\boldsymbol{\alpha}_g$) while injecting new semantic information ($\boldsymbol{\beta}_g$).

Formally, consider a generic feature group $g$ with original tokens $\mathbf{X}_g^{(l)} \in \mathbb{R}^{N_g \times d}$ and its updated hubs $\tilde{\mathbf{X}}_g^{(l+1)}$. The BGU operates in two steps: \textit{Channel-wise Parameter Generation} and \textit{Token-wise Broadcasting}.

First, we flatten the updated hub representations to capture the global configuration of the group and project them into $\mathbb{R}^{2d}$ to generate two modulation vectors: the scaling vector $\boldsymbol{\alpha}_g \in \mathbb{R}^d$ and the shifting vector $\boldsymbol{\beta}_g \in \mathbb{R}^d$:
\begin{equation}
    \boldsymbol{\alpha}_g, \boldsymbol{\beta}_g = \operatorname{Split}\left( \operatorname{MLP}_{\mathrm{bgu}}\bigl(\operatorname{Flatten}(\tilde{\mathbf{X}}_g^{(l+1)})\bigr) \right).
\end{equation}
Here, $\boldsymbol{\alpha}_g$ and $\boldsymbol{\beta}_g$ represent the learned global context. 

Second, we broadcast these parameters to every token in the group. For each token $\mathbf{x}_{i}$ in $\mathbf{X}_g^{(l)}$ (where $i=1,\dots,N_g$), the modulation is applied as:
\begin{equation}
    \operatorname{BGU}(\mathbf{x}_{i}, \tilde{\mathbf{X}}_g^{(l+1)}) = \mathbf{x}_{i} \odot \sigma(\boldsymbol{\alpha}_g) + \boldsymbol{\beta}_g.
\end{equation}
where $\sigma(\cdot)$ denotes the sigmoid activation function. By sharing $\boldsymbol{\alpha}_g$ and $\boldsymbol{\beta}_g$ across all $N_g$ tokens, BGU effectively performs a token-wise broadcast, ensuring that the group-level consensus uniformly recalibrates the local feature channels without quadratic computational cost.

\noindent \textbf{Unified Feature Refinement.}
With the generic BGU defined, the refinement for all three groups follows a standardized residual protocol. For any group $g \in \{C, S, T\}$, the original tokens for the next layer are updated as:
\begin{equation}
    \mathbf{X}_g^{(l+1)} = \operatorname{LN}\Bigl( \mathbf{X}_g^{(l)} + \operatorname{BGU}(\mathbf{X}_g^{(l)}, \tilde{\mathbf{X}}_g^{(l+1)}) \Bigr).
\end{equation}
This unified formulation ensures that categorical features, behavior sequences, and task tokens are all iteratively refined by their respective global contexts. Crucially, since the broadcasting operation is strictly width-preserving, the output $\mathbf{X}_g^{(l+1)}$ maintains the exact same dimensions as the input $\mathbf{X}_g^{(l)}$. This design allows straightforward stacking of multiple INFNet blocks without the need for pooling or upsampling operations.

\noindent \textbf{Complexity Analysis.} 
The computational cost of Phase 2 is composed of two parts: parameter generation and broadcast application. 
The parameter generation step processes the flattened hubs, incurring a complexity of $O(N_{\mathrm{hub}} d^2)$, which depends solely on the number of hubs and is independent of the input token scale.
The broadcast step applies element-wise operations to the original tokens, scaling linearly as $O(N_{\mathrm{in}} d)$.
Combining Phase 1 ($O(N_{\mathrm{hub}} N_{\mathrm{in}} d)$) and Phase 2, the total complexity of an INFNet block is $O(N_{\mathrm{in}} d (N_{\mathrm{hub}} + 1) + N_{\mathrm{hub}} d^2)$. Since the number of hubs is fixed and small ($N_{\mathrm{hub}} \ll N_{\mathrm{in}}$), the overall complexity remains strictly linear $O(N_{\mathrm{in}})$, ensuring efficiency comparable to simple MLP interactions but with deeper expressiveness.

\subsection{Multi-Task Optimization}
After stacking $L$ INFNet blocks, we obtain the final task-specific representations $\mathbf{T}^{(L)} = \{\mathbf{t}_1^{(L)}, \dots, \mathbf{t}_{N_t}^{(L)}\}$, which have been iteratively refined by both global feature context and shared task knowledge. The prediction probability $\hat{y}_i$ for task $i$ is generated by a task-specific MLP head followed by a sigmoid activation:
\begin{equation}
    \hat{y}_i = \sigma\left( \operatorname{MLP}_{i}(\mathbf{t}_i^{(L)}) \right).
\end{equation}
The entire network is optimized end-to-end using a weighted binary cross-entropy loss:
\begin{equation}
    \mathcal{L} = \sum_{i=1}^{N_t} \lambda_i \mathcal{L}_{\mathrm{BCE}}(\hat{y}_i, y_i),
\end{equation}
where $y_i$ denotes the ground-truth label and $\lambda_i$ balances the importance of different objectives.

\begin{table}[b]
  \centering
  \caption{Statistics of the datasets used in our experiments.}
  \label{tab:dataset}
  \scalebox{0.95}{
    \begin{tabular}{lcc}
    \toprule
    \textbf{Dataset} & \textbf{Industrial} & \textbf{KuaiRand} \\
    \midrule
    \# Users & $>$10M & 27K \\
    \# Items & $>$1M & 32M \\
    \# Interactions & $>$1B & 322M \\
    \# Categorical Fields & $>$100 & 89 \\
    \# Sequence Fields & $>$100 & 28 \\
    \# Tasks & 5 & 3 \\
    \bottomrule
    \end{tabular}}
\end{table}

%% file: 4.exp.tex
\label{sec:experiments}

\begin{table*}[t]
\centering
\setlength{\tabcolsep}{1.3pt} 
\renewcommand{\arraystretch}{1.2} 
\scriptsize
\caption{Detailed performance comparison across all individual tasks on \textbf{KuaiRand} and \textbf{Industrial} datasets. We report model parameters (\textbf{Params}) and computational complexity (\textbf{FLOPs}). \baby{} achieves SOTA performance with significantly lower cost. The best results are \textbf{bolded}, second-best are \underline{underlined}.}
\label{tab:all_detailed_tasks_final}
\resizebox{\textwidth}{!}{%
\begin{tabular}{ll|cc|cccccc|cc|cccccccccc}
\toprule
\multirow{3}{*}{\textbf{Category}} & \multirow{3}{*}{\textbf{Model}} & \multicolumn{8}{c|}{\textbf{KuaiRand (3 Tasks)}} & \multicolumn{12}{c}{\textbf{Industrial Dataset (5 Tasks)}} \\
\cmidrule(lr){3-10} \cmidrule(lr){11-22}
 & & \multicolumn{2}{c|}{\textbf{Efficiency}} & \multicolumn{2}{c}{Click} & \multicolumn{2}{c}{Like} & \multicolumn{2}{c|}{Long-View} & \multicolumn{2}{c|}{\textbf{Efficiency}} & \multicolumn{2}{c}{Click} & \multicolumn{2}{c}{Play 3s} & \multicolumn{2}{c}{Play 5s} & \multicolumn{2}{c}{Play End} & \multicolumn{2}{c}{Follow} \\
\cmidrule(lr){3-4} \cmidrule(lr){5-6} \cmidrule(lr){7-8} \cmidrule(lr){9-10} \cmidrule(lr){11-12} \cmidrule(lr){13-14} \cmidrule(lr){15-16} \cmidrule(lr){17-18} \cmidrule(lr){19-20} \cmidrule(lr){21-22}
 & & \tiny Params & \tiny FLOPs & \tiny AUC & \tiny GAUC & \tiny AUC & \tiny GAUC & \tiny AUC & \tiny GAUC & \tiny Params & \tiny FLOPs & \tiny AUC & \tiny GAUC & \tiny AUC & \tiny GAUC & \tiny AUC & \tiny GAUC & \tiny AUC & \tiny GAUC & \tiny AUC & \tiny GAUC \\
\midrule
\multirow{3}{*}{\textit{Foundations}} 
 & DeepFM & 3.0M & 12.9G & 0.7615 & 0.6540 & 0.8752 & 0.6110 & 0.7584 & 0.6550 & 95M & 410G & 0.9225 & 0.7250 & 0.8490 & 0.6900 & 0.8525 & 0.6810 & 0.8950 & 0.7680 & 0.9665 & 0.6820 \\
 & DCNv2 & 3.5M & 22.7G & 0.7632 & 0.6569 & 0.8784 & 0.6154 & 0.7598 & 0.6569 & 95M & 615G & 0.9248 & 0.7282 & 0.8515 & 0.6935 & 0.8552 & 0.6845 & 0.8972 & 0.7715 & 0.9688 & 0.6852 \\
 & AutoInt & 4.0M & 33.5G & 0.7628 & 0.6550 & 0.8775 & 0.6130 & 0.7590 & 0.6550 & 98M & 820G & 0.9242 & 0.7270 & 0.8505 & 0.6920 & 0.8545 & 0.6830 & 0.8965 & 0.7702 & 0.9680 & 0.6840 \\
\midrule
\multirow{3}{*}{\textit{Sequence}} 
 & DIN & 2.1M & 31.3G & 0.7596 & 0.6470 & 0.8745 & 0.6050 & 0.7572 & 0.6530 & 96M & 1433G & 0.9262 & 0.7310 & 0.8542 & 0.6985 & 0.8582 & 0.6902 & 0.9002 & 0.7782 & 0.9712 & 0.6885 \\
 & BST & 3.2M & 149.8G & 0.7652 & 0.6585 & 0.8840 & 0.6220 & 0.7635 & 0.6610 & 105M & 4915G & 0.9285 & 0.7345 & 0.8575 & 0.7025 & 0.8615 & 0.6942 & 0.9028 & 0.7818 & 0.9735 & 0.6932 \\
 & HSTU & 1.5M & 85.3G & 0.7674 & 0.6602 & 0.8885 & 0.6285 & 0.7668 & 0.6642 & 108M & 6144G & 0.9308 & 0.7382 & 0.8595 & 0.7052 & 0.8632 & 0.6975 & 0.9045 & 0.7842 & 0.9752 & 0.6965 \\
\midrule
\multirow{3}{*}{\textit{Efficient}} 
 & WuKong & 1.8M & 6.02G & 0.7668 & 0.6604 & 0.8872 & 0.6275 & 0.7657 & 0.6635 & 98M & 328G & 0.9290 & 0.7338 & 0.8585 & 0.7042 & 0.8622 & 0.6965 & 0.9030 & 0.7825 & 0.9745 & 0.6952 \\
 & RankMixer & 2.0M & 9.63G & 0.7670 & 0.6612 & 0.8880 & 0.6292 & 0.7662 & 0.6648 & 102M & 491G & 0.9302 & 0.7355 & 0.8602 & 0.7065 & 0.8640 & 0.6988 & 0.9048 & 0.7852 & 0.9760 & 0.6982 \\
 & OneTrans & 1.6M & 6.56G & 0.7665 & 0.6608 & 0.8868 & 0.6270 & 0.7655 & 0.6638 & 100M & 410G & 0.9298 & 0.7342 & 0.8592 & 0.7052 & 0.8632 & 0.6975 & 0.9038 & 0.7840 & 0.9752 & 0.6970 \\
\midrule
\multirow{2}{*}{\textit{Composite}} 
 & RM+PLE & 2.5M & 22.9G & 0.7682 & 0.6628 & \underline{0.8898} & 0.6315 & 0.7672 & 0.6655 & 112M & 1024G & 0.9324 & 0.7412 & 0.8608 & 0.7085 & 0.8645 & 0.7005 & 0.9062 & 0.7878 & 0.9772 & 0.7005 \\
 & OT+PLE & 2.1M & 15.7G & \underline{0.7685} & \underline{0.6634} & 0.8890 & \underline{0.6322} & \underline{0.7685} & \underline{0.6662} & 110M & 820G & \underline{0.9332} & \underline{0.7425} & \underline{0.8615} & \underline{0.7102} & \underline{0.8652} & \underline{0.7018} & \underline{0.9029} & \underline{0.7895} & \underline{0.9760} & \underline{0.7025} \\
\midrule
\textbf{Ours} & \textbf{INFNet} & 1.6M & 3.24G & \textbf{0.7736} & \textbf{0.6757} & \textbf{0.8960} & \textbf{0.6464} & \textbf{0.7727} & \textbf{0.6784} & 100M & 
 202G & \textbf{0.9381} & \textbf{0.7492} & \textbf{0.8694} & \textbf{0.7180} & \textbf{0.8744} & \textbf{0.7125} & \textbf{0.9092} & \textbf{0.7966} & \textbf{0.9790} & \textbf{0.7052} \\
\midrule
\multicolumn{2}{l|}{\textit{Improv. vs. Best}} & - & - & +0.66\% & +1.85\% & +0.79\% & +2.25\% & +0.55\% & +1.83\% & - & - & +0.52\% & +0.90\% & +0.92\% & +1.10\% & +1.06\% & +1.52\% & +0.70\% & +0.90\% & +0.31\% & +0.38\% \\
\bottomrule
\end{tabular}%
}
\end{table*}

In this section, we rigorously evaluate the performance of \baby by addressing the following research questions:
\begin{itemize}[leftmargin=*]
    \item \textbf{RQ1:} Does \baby outperform state-of-the-art baselines, particularly compared to heavy sequence models (e.g., HSTU) and efficient unified architectures (e.g., OneTrans, RankMixer), across diverse datasets?
    \item \textbf{RQ2:} Are the proposed core components (e.g., Hub-based interaction, Task Tokens, BGU) essential for model performance? Does the model learn meaningful task-aware patterns as designed?
    \item \textbf{RQ3:} Can \baby achieve a superior trade-off between inference accuracy and computational cost (FLOPs/Params)? How does the model scale with varying hub configurations?
    \item \textbf{RQ4:} How does \baby perform in a real-world, large-scale online recommendation system in terms of multi-object, and serving latency?
\end{itemize}

To answer \textbf{RQ1}, we present a comprehensive performance comparison across three benchmarks in Section~\ref{sec:overall_performance}. 
For \textbf{RQ2}, we conduct extensive macro and micro ablation studies in Section~\ref{sec:ablation} and visualize the attention patterns in Section~\ref{sec:visualization}. 
To address \textbf{RQ3}, we analyze the hyperparameter sensitivity and scaling laws to demonstrate the efficiency advantages in Section~\ref{sec:efficiency}. 
Finally, for \textbf{RQ4}, we report the results of rigorous online A/B testing in Section~\ref{sec:online_ab}.

\subsection{Experimental Setup}

\paratitle{Datasets.}
We conduct comprehensive evaluations of \baby{} on two benchmarks varying in scale and complexity, as summarized in Table~\ref{tab:dataset}:
\begin{itemize}[leftmargin=*]
    \item \textbf{KuaiRand}~\cite{gao2022kuairand}: A public recommendation dataset containing three interaction tasks (Click, Like, Long-view). Given its representative nature in open-source benchmarks, we utilize it primarily for architectural validation and ablation studies.

    \item \textbf{Industrial}: A large-scale proprietary dataset collected from a real-world short-video production environment, comprising over 1 billion interactions and five optimization objectives (Click, Play 3s, Play 5s, Play End, Follow). With its extremely high feature dimensionality, this dataset serves as the critical testbed for analyzing the scalability and robustness of \baby{} in real-world scenarios.
\end{itemize}

\paratitle{Baselines.}
We compare \baby against 11 strong baselines, categorized into four groups to highlight different competitive angles.

\noindent \textbf{Foundations.}
These models serve as the fundamental benchmarks for capturing high-order feature interactions.
\begin{itemize}[leftmargin=1.5em, label=$\bullet$]
    \item \textbf{DeepFM}~\cite{guo2017deepfm}: A classic architecture that combines FM (Factorization Machines) for low-order interactions and deep neural networks for high-order feature learning.
    \item \textbf{DCNv2}~\cite{wang2021dcn}: An enhanced Deep \& Cross Network that explicitly models feature crossing at vector-wise level to learn predictive interactions.
    \item \textbf{AutoInt}~\cite{song2019autoint}: Utilizes a multi-head self-attention network to automatically learn high-order feature interactions in an explicit fashion.
\end{itemize}

\noindent \textbf{Sequence Modeling.}
 These methods employ advanced mechanisms like target-attention or Transformers to model sequential dependencies. 
\begin{itemize}[leftmargin=1.5em, label=$\bullet$]
    \item \textbf{DIN}~\cite{zhou2018deep}: Introduces a target-attention mechanism to activate historical behaviors relevant to the candidate item.
    \item \textbf{BST}~\cite{chen2019behavior}: Adapts the Transformer architecture to user behavior sequences, capturing long-term dependencies via self-attention.
    \item \textbf{HSTU}~\cite{zhai2024actions}: A recent high-performance Transformer-based user action model optimized for large-scale retrieval and ranking.
\end{itemize}

\noindent \textbf{Efficient SOTA.}
These state-of-the-art architectures aim to replace heavy canonical attention mechanisms with more efficient operations while demonstrating robust scalability.
\begin{itemize}[leftmargin=1.5em, label=$\bullet$]
    \item \textbf{WuKong}~\cite{zhang2024wukong}: Adopts a Factorized Interaction mechanism to capture high-order feature interactions efficiently via low-rank matrix decomposition.
    \item \textbf{RankMixer}~\cite{zhu2025rankmixer}: Design a novel Token Mixer to facilitate feature interaction and employs a Mixture-of-Experts (MoE) routing strategy, thereby demonstrating superior scalability and capacity.
    \item \textbf{OneTrans}~\cite{zhang2025onetrans}: Proposes a Unified Transformer framework that seamlessly integrates both sequential user behaviors and non-sequential context features into a single modeling stream, simplifying the architecture while maintaining effectiveness.
\end{itemize}

\noindent \textbf{Composite Baselines.}
To evaluate multi-task performance fairly, we combine efficient backbones with the PLE\cite{tang2020progressive} multi-task head.
\begin{itemize}[leftmargin=1.5em, label=$\bullet$]
    \item \textbf{RankMixer + PLE(RM+PLE)}: Combines the RankMixer backbone with (PLE)~\cite{tang2020progressive} for multi-task learning.
    \item \textbf{OneTrans + PLE(OT+PLE)}: Extends OneTrans with PLE to handle multiple optimization objectives simultaneously.
\end{itemize}

\paratitle{Metrics \& Experimental Settings.}
Following \cite{zhu2025rankmixer}, we report two standard metrics: (1) \textbf{AUC} (area under the receiver operating characteristic curve) measures overall ranking quality higher is better; and (2) \textbf{GAUC}, a user-level averaged AUC that weights users uniformly to assess personalization quality higher is better.

To ensure a fair comparison, we align the data pipeline and evaluation protocols across all experiments, though the implementation frameworks differ by dataset scale. 
For the public datasets KuaiRand, we utilize the open-source benchmark library \texttt{FuxiCTR}\footnote{\url{https://github.com/reczoo/FuxiCTR}}~\cite{zhu2021open} with a standard 8:1:1 random split for training, validation, and testing.
For the industrial dataset, models are implemented in TensorFlow and evaluated using a strict time-series split to simulate production environments. Following \cite{zhu2025rankmixer,zhang2025onetrans}, we fix the batch size at $4096$ and grid-search the embedding dimensions in $\{8, \dots, 128\}$ and learning rates in $\{1\mathrm{e}{-4}, \dots, 1\mathrm{e}{-2}\}$ with Adam or Adagrad optimizers. All experiments are conducted on NVIDIA A100 GPUs, with average results from 5 independent runs reported to ensure reproducibility.

Unless otherwise specified, for baselines originally designed for single-task ranking (e.g., DeepFM, DIN, and OneTrans), we adapt them to the multi-task setting using the standard Shared-Bottom architecture. Specifically, the model backbone serves as a shared encoder, followed by $N_t$ independent task-specific towers (implemented as 2-layer MLPs) to generate predictions for all tasks simultaneously. Baselines with explicit multi-task designs (e.g., HSTU and RankMixer variants) follow their original architectures.

\subsection{Overall Performance Comparison (RQ1)} \label{sec:overall_performance}

Table~\ref{tab:all_detailed_tasks_final} summarizes the overall performance comparison on the KuaiRand and Industrial datasets, reporting per-task AUC/GAUC together with model size and FLOPs. Across both datasets and all tasks, INFNet consistently achieves the top performance. In terms of accuracy, the improvements over the strongest baselines are stable and non-trivial. On KuaiRand, INFNet delivers clear AUC gains across all three objectives. On the Industrial dataset, INFNet improves AUC over the best baseline by a consistent margin across five objectives, with the largest gains appearing on click and early-retention signals and positive gains also observed on deeper-retention and follow signals. Specifically, \baby{} achieves a 0.005 lift in Click AUC on the Industrial dataset. In the context of large-scale online recommendation systems, where baselines are already heavily optimized, a five-mille improvement is considered a substantial breakthrough that typically translates into significant revenue growth and enhanced user engagement.

Beyond effectiveness, INFNet achieves these gains with a favorable efficiency profile. Compared with heavy sequence models, INFNet attains better accuracy under substantially lower FLOPs, indicating that the proposed aggregate-and-broadcast interaction provides a more compute-efficient path to deep cross-feature modeling than dense token-wise interaction. Compared with efficient baselines, INFNet maintains a similar or lower computational footprint while improving accuracy across tasks, suggesting that its interaction mechanism captures additional cross-type and task-conditioned signals without sacrificing serving efficiency. Finally, INFNet also outperforms composite baselines that combine an efficient interaction backbone with a PLE-style multi-task module, suggesting that injecting task signals into the interaction backbone via task tokens provides additional benefits beyond adding a multi-task module on top.

\subsection{Ablation Studies (RQ2)}\label{sec:ablation}
To rigorously validate the effectiveness of \baby{}, we conduct a comprehensive ablation analysis on the large-scale industrial dataset.

\subsubsection{Component Effectiveness (Macro Analysis)}
\begin{figure}[t]
    \centering
    \includegraphics[width=1\linewidth]{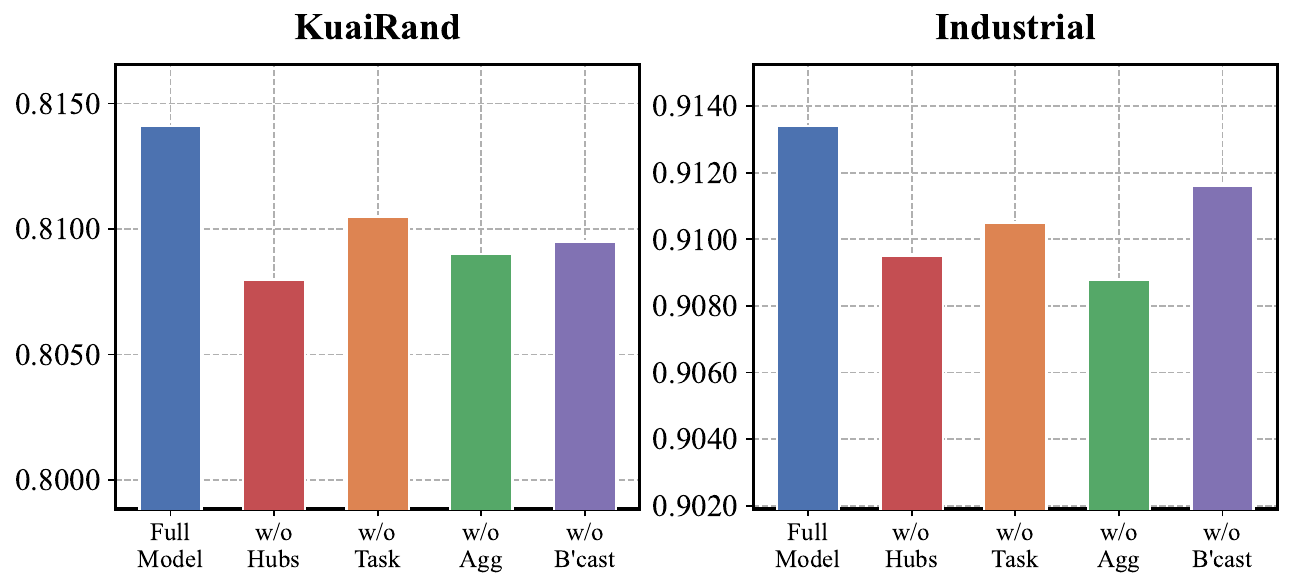}
    \caption{\small{Component effectiveness analysis. We report the average AUC across all tasks for \baby{} and its variants, each representing the removal or degradation of a key interaction phase.}}
    \label{fig:ablation_macro}
\end{figure}

To verify the core pillars of \baby, we evaluate four structural regressions calibrated to maintain similar parameter counts and complexity as the full model (Fig.~\ref{fig:ablation_macro}). First, the \textit{w/o Hubs} variant suffers from significant degradation, demonstrating that all-to-all attention lacks the necessary architectural constraints to scale effectively for multi-task modeling, whereas \baby{} maintains high signal fidelity through its structured bottleneck. Second, removing task tokens (\textit{w/o Task}) forces a task-blind, Shared-Bottom architecture, proving that explicit task guidance is prerequisite for mitigating negative transfer. Finally, we dissect the two-phase interaction loop: disabling Phase 1 cross-view aggregation (\textit{w/o Agg}) degrades the model into a Late Fusion paradigm with the sharpest performance decline, while disabling Phase 2 (\textit{w/o B'cast}) results in a Latent-Only model. The latter's performance remains inferior to any sub-optimal implementation in our micro-analysis, reinforcing that the feedback loop for re-injecting global consensus into local features is critical for final accuracy.

\subsubsection{Fine-grained Design Analysis (Micro Analysis)}
Beyond macro-architectural components, implementation details significantly influence \baby's effectiveness. As shown in Table~\ref{tab:micro_ablation}, we investigate structural choices across five dimensions on the Industrial dataset. For hub initialization, our MLP Projection and Type-Specific Aggregation outperform random or pooling-based baselines, confirming that learnable, non-linear mappings and semantic granularity are essential for compressing heterogeneous fields and behaviors. Regarding interaction, the Concat-Linear fusion and Hybrid task hub construction yield superior results over simplistic summation or single-token designs, validating the necessity of modeling both task-specific nuances and shared cross-view knowledge. Finally, the ablation of the Broadcast Gated Unit (BGU) demonstrates that the Affine transformation is more effective than either pure scaling or shifting, as it enables simultaneous feature modulation and global context injection. These consistent, albeit subtle, performance gains across all variants reinforce the optimality of our refined structural designs.

\begin{figure}[b]
    \centering
    \includegraphics[width=1\linewidth]{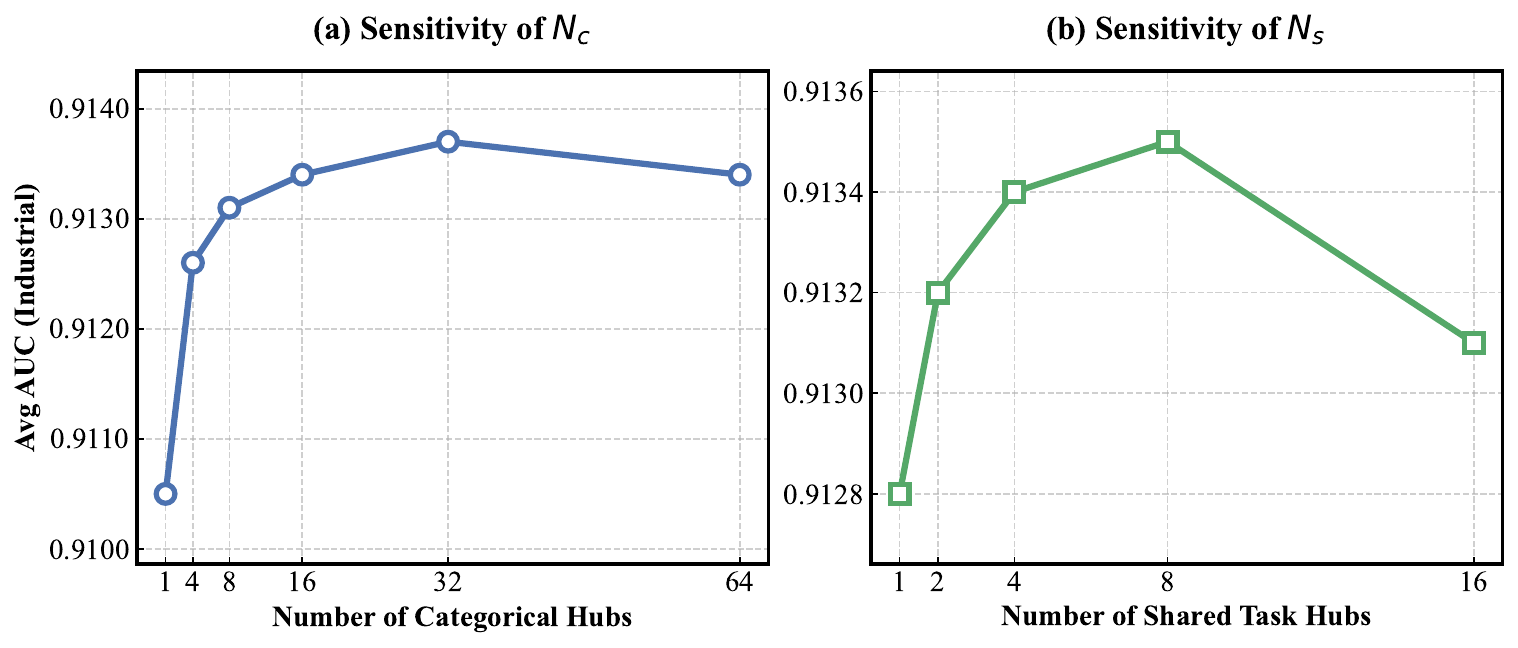}
    \caption{\small{Sensitivity analysis of hub composition. We evaluate the impact of (a) the number of categorical hubs $N_c$ and (b) shared task hubs $N_s$ on the Industrial dataset.}}
    \label{fig:hub_sensitivity}
\end{figure}

\subsection{Efficiency and Scalability Analysis (RQ3)}\label{sec:efficiency}

To further demonstrate the superiority of \baby over efficient SOTA architectures, we conduct an efficiency analysis on the industrial dataset.

\begin{figure*}[h] \centering \includegraphics[width=1.0\linewidth]{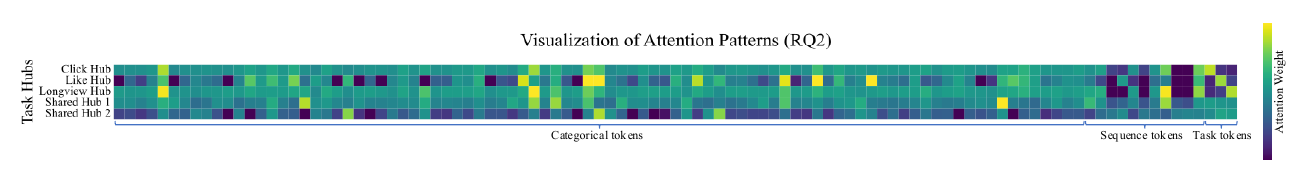} \caption{\small{Heatmap of cross-attention weights between task hubs (Y-axis) and input tokens (X-axis) in KuaiRand dataset. For visualization purposes, extensive sequence tokens are aggregated by their behavior types.}} \label{fig:visual}
\end{figure*}

\subsubsection{Hyperparameter Sensitivity: Hub Composition}
We investigate \baby's sensitivity to the number of categorical hubs ($N_c$) and shared task hubs ($N_s$) on the Industrial dataset (Fig.~\ref{fig:hub_sensitivity}). For categorical features, performance significantly improves as $N_c$ increases from 1 and peaks at $N_c=32$, confirming that while the structured bottleneck is essential, sufficient latent capacity is required to resolve complex co-occurrence patterns across hundreds of fields. Notably, even the extreme bottleneck of $N_c=1$ outperforms the \textit{w/o Agg} variant, demonstrating the inherent effectiveness of the aggregate-broadcast mechanism. Regarding multi-task coordination, the model achieves optimal results at $N_s=4$; providing even a single shared hub ($N_s=1$) yields better results than the \textit{Specific Only} micro-variant, validating the value of modeling cross-task consensus. While $N_c=32$ offers the peak performance, we select $N_c=16$ and $N_s=4$ as default parameters to achieve an optimal trade-off between predictive accuracy and computational efficiency.

\begin{table}[t]
\centering
\caption{\small{Fine-grained analysis of structural design choices on the Industrial dataset. We report the Avg AUC across five tasks. The default settings of \baby{} are indicated in \textbf{bold}.}}
\label{tab:micro_ablation}
\renewcommand{\arraystretch}{1.15} 
\setlength{\tabcolsep}{4pt} 
\resizebox{1.0\linewidth}{!}{%
\begin{tabular}{c|l|cc}
\toprule
\textbf{Design Aspect} & \textbf{Variant} & \textbf{Avg AUC} $\uparrow$ & \textbf{Rel. Drop ($\Delta$)} \\
\midrule
\multirow{3}{*}{\shortstack{\textbf{Categorical Hub}\\\textbf{Initialization}}} 
 & Random Initialization & 0.9115 & -0.21\% \\
 & Mean Pooling (Fields) & 0.9124 & -0.11\% \\
 & \textbf{MLP Projection (Ours)} & \textbf{0.9134} & \textbf{--} \\
\midrule
\multirow{3}{*}{\shortstack{\textbf{Sequence Hub}\\\textbf{Initialization}}} 
 & Random Initialization & 0.9112 & -0.24\% \\
 & Global Mean Pooling & 0.9120 & -0.15\% \\
 & \textbf{Type-Specific Agg. (Ours)} & \textbf{0.9134} & \textbf{--} \\
\midrule
\multirow{3}{*}{\shortstack{\textbf{Multi-View}\\\textbf{Fusion}}} 
 & Summation & 0.9128 & -0.06\% \\
 & Self-Attn + Concat-Linear & 0.9130 & -0.04\% \\
 & \textbf{Concat-Linear (Ours)} & \textbf{0.9134} & \textbf{--} \\
\midrule
\multirow{3}{*}{\shortstack{\textbf{Task Hub}\\\textbf{Construction}}} 
 & Specific Only ($\mathbf{T}$) & 0.9126 & -0.09\% \\
 & Shared Only ($\tilde{\mathbf{T}}_{\mathrm{shared}}$) & 0.9122 & -0.13\% \\
 & \textbf{Hybrid (Ours)} & \textbf{0.9134} & \textbf{--} \\
\midrule
\multirow{3}{*}{\shortstack{\textbf{Broadcast Unit}\\\textbf{(BGU)}}} 
 & Only Shifting ($\mathbf{x} + \boldsymbol{\beta}$) & 0.9123 & -0.11\% \\
 & Only Scaling ($\mathbf{x} \odot \boldsymbol \sigma({\alpha})$) & 0.9128 & -0.06\% \\
 & \textbf{Affine ($\mathbf{x} \odot \boldsymbol{\sigma(\alpha)} + \boldsymbol{\beta}$)} & \textbf{0.9134} & \textbf{--} \\
\bottomrule
\end{tabular}%
}
\end{table}

\begin{figure}[t]
    \centering
    \includegraphics[width=1.0\linewidth]{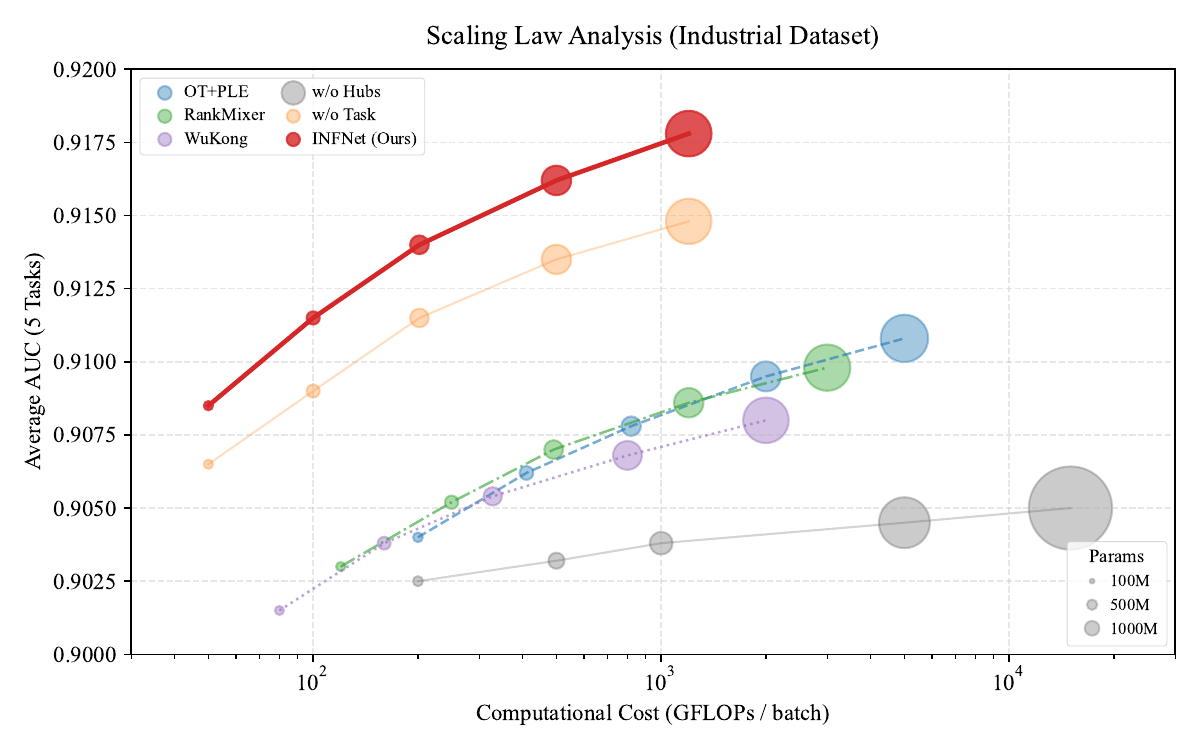}
    \caption{\small{Scaling Laws and Mechanism Ablation. We compare the scaling trajectories of the full \baby{} against efficient baselines and its key ablated variants.}}
    \label{fig:scaling_law}
\end{figure}

\subsubsection{Efficiency and Scalability Analysis}
We evaluate the scalability of \baby{} on the Industrial dataset by comparing it against efficient baselines and its own ablated variants across a wide range of computational budgets. 

As shown in Figure~\ref{fig:scaling_law}, \baby{} exhibits a significantly steeper scaling slope than specialized baselines (e.g., WuKong, OT+PLE), demonstrating its superior efficiency in translating increased FLOPs into predictive power. Crucially, we observe consistent performance gains when scaling along both network depth and embedding width dimensions. Based on these results, we identify an embedding dimension of 64 and a 4-layer stack (100M params) as the optimal balance between capacity and industrial latency constraints.

Ablation studies further reveal two critical insights: 
(1) The performance gap between \baby{} and the \textit{w/o Task Tokens} variant widens as FLOPs increase, proving that task-aware routing is essential for effectively utilizing higher model capacity. 
(2) The \textit{w/o Hubs} variant displays a nearly flat scaling curve, failing to improve significantly even with higher resource allocation. This confirms that our hub-mediated mechanism is the fundamental driver of signal fidelity and parameter efficiency in complex multi-task environments.

\subsection{Visualization of Attention Patterns (RQ2)}\label{sec:visualization}
To interpret how task tokens facilitate multi-task modeling, we visualize the cross-attention weights in the final block in KuaiRand dataset. As shown in Figure~\ref{fig:visual}, the Y-axis represents different task hubs (specific and shared), while the X-axis denotes all input tokens, including categorical fields, behavior sequences (summed for clarity), and task identifiers.

The visualization reveals three key insights. First, there is a distinct functional divergence: shared hubs exhibit a smoother and more distributed attention pattern compared to the specific hubs. This suggests that shared tokens aggregate global context and common interests, while specific hubs act as sharp filters for task-relevant signals. Second, different tasks demonstrate unique dependencies on feature groups; for instance, the click task may rely more on categorical profiles, while the follow task focuses intensely on specific long-term behaviors. Finally, the heterogeneous attention bands across task rows confirm that the model effectively disentangles conflicting objectives. By allowing tasks to attend to the same feature space with varying intensities and preferences, \baby{} successfully captures both task-specific nuances and shared cross-task correlations.

\subsection{Online A/B Testing (RQ4)}\label{sec:online_ab}

We deployed \baby in a large-scale online short-video advertising platform serving billions of requests per day. The model was trained in a streaming setting and evaluated against a production baseline (approximately OneTrans + PLE) through a rigorous one-month A/B test.

\begin{table}[h]
\centering
\caption{\small{Online A/B test results. \baby achieves significant gains in both revenue and user engagement metrics with lower latency.}}
\label{tab:abtest}
\scalebox{0.9}{
\begin{tabular}{lcccccc}
\toprule
Method & REV & CTR & P3s & P5s & PEnd & Latency \\
\midrule
Baseline & -- & -- & -- & -- & -- & 18.28ms \\
\baby  & \textbf{+1.59\%} & \textbf{+1.16\%} & \textbf{+0.11\%} & \textbf{+0.32\%} & \textbf{+0.35\%} & \textbf{18.17ms} \\
\bottomrule
\end{tabular}}
\end{table}

As shown in Table~\ref{tab:abtest}, \baby achieved a \textbf{+1.587\%} increase in Revenue and \textbf{+1.155\%} in CTR, while maintaining essentially unchanged prediction latency (within \textbf{0.11ms}). The consistent improvement in retention metrics (P3s, P5s, PEnd) indicates that \baby enhances user engagement alongside platform revenue, demonstrating a better balance across multiple objectives in real-world ranking.

%% file: 5.con.tex
This paper identifies "early aggregation" and "late fusion" as fundamental information-flow bottlenecks in efficient ranking. We propose \baby, a width-preserving framework that utilizes a hub-mediated \textit{aggregate-and-broadcast} mechanism to maintain signal fidelity with linear complexity. Experiments confirm \baby's superior scaling capacity and online success further validates \baby as a scalable foundational architecture for complex industrial multi-task learning.